\documentclass[twocolumn]{aastex631}
\usepackage{amsmath,amssymb,graphicx,longtable,booktabs,xcolor}
\usepackage{commath}
\usepackage{floatrow}
\usepackage{soul}
\usepackage{CJK}
\usepackage{threeparttable}
\usepackage{soul}
\usepackage{ulem}

\def\msyr{~$M_{\odot}$~yr$^{-1}$}
\def\mspc{~$M_{\odot}$~pc$^{-2}$}
\def\mspcgyr{~$M_{\odot}$~pc$^{-2}$~Gyr$^{-1}$}

\def\hi{H~{\sc i}~}

\newcommand{\ms}{$M_{\odot}$}

\shorttitle{SFH-regulated Chemical Evolution Models}
\shortauthors{Yin et al.}
\begin{document}

\title{Linking the Metallicity Enrichment History to the Star Formation History: An SFH-regulated Chemical Evolution Model and Its Implications for the Gas Cycling Process}

\author[0000-0002-4499-1956]{Jun Yin}
\affiliation{Key Laboratory for Research in Galaxies and Cosmology, Shanghai Astronomical Observatory, Chinese Academy of Sciences, 80 Nandan Road, Shanghai 200030, People's Republic of China}
\affiliation{Key Lab for Astrophysics, Shanghai, 200034, People's Republic of China}

\author[0000-0002-3073-5871]{Shiyin Shen}
\affiliation{Key Laboratory for Research in Galaxies and Cosmology, Shanghai Astronomical Observatory, Chinese Academy of Sciences, 80 Nandan Road, Shanghai 200030, People's Republic of China}
\affiliation{Key Lab for Astrophysics, Shanghai, 200034, People's Republic of China}

\author[0000-0003-2478-9723]{Lei Hao}
\affiliation{Key Laboratory for Research in Galaxies and Cosmology, Shanghai Astronomical Observatory, Chinese Academy of Sciences, 80 Nandan Road, Shanghai 200030, People's Republic of China}

\correspondingauthor{Jun Yin}
\email{jyin@shao.ac.cn}

%--------------------------------------------------------------------
% abstract
\begin{abstract}

The metallicity enrichment history (MEH) of a galaxy is determined by its star formation history (SFH) and the gas cycling process. In this paper, we construct a chemical evolution model that is regulated by the SFH of the system. In this SFH-regulated model, the evolution of all other variables, including the MEH, can be determined by the SFH. We test this model on six locally isolated dwarf galaxies covering three dwarf types that were observed by the Local Cosmology from Isolated Dwarfs (LCID) project. The SFHs and MEHs of these LCID galaxies have been measured from the deep color-magnitude diagrams that are down to the main sequence turn-offs stars. With simple assumptions of the star formation law and the mass-dependent outflows, our SFH-regulated model successfully reproduces the MEHs of all six LCID galaxies from their SFHs, with only one free parameter, the wind efficiency $\eta \sim 1.0$, for all six galaxies. This model provides a physically motivated link that directly connects the SFH and MEH of a galaxy, which will be useful to accommodate into the state-of-the-art stellar population synthesis models to help relieve the nuisance of the heavy degeneracy between the ages and metallicities of the stellar populations.

\end{abstract}


%--------------------------------------------------------------------
% main body

\section{Introduction}
\label{sec:intro}

Gas cycling plays an important role in the evolution history of a galaxy. The primordial gas with low metallicity cools down from the halo and feeds the formation of stars, whereas the evolved stars blow the metal-enriched gas out into the surrounding interstellar medium (ISM) or even further out to the halo or the intergalactic medium via outflows driven by supernovae (SNe), stellar winds, and/or radiative pressure \citep{naab17}. Among this baryonic cycle, there are two key observables tightly linked to the processes of how the gas is cooled, accreted, consumed, and blown out. One is the star-formation history (SFH) of a galaxy, i.e., the number of stars formed in a galaxy as a function of time. Another is the metal content of the ISM, which could be enriched by an SN explosion or a stellar wind carried out by the outflow or diluted by the gas inflow. Therefore, the combination of the SFH and the metallicity enrichment history (MEH) provides us with a window to look into the gas cycling process and the evolution of galaxies.

The chemical evolution model (CEM), which links the MEH of a galaxy with its gas cycling and star formation (SF) process, has been developed for decades and is widely used in the study of nearby galaxies \citep[e.g.,][]{schmidt63,lynden75,Pagel75,Carigi94,Prantzos08,yin09,yin10,yin11,chang12,dave12,kang12,lilly13,molla15,zhu17,barrera18,maiolino19,matteucci21}.
Complex gas cycling and SF processes are typically simplified with a number of analytical formulae. 
In particular, it is assumed that the SFH is related to the gas content in the reservoir following the SF law, and the inferred present-day metallicity has been compared with the observations to constrain the model parameters (see Section 2 for details). However, when considering the detailed physical processes in the CEMs, we note that the constraining power of the ``present-day metallicity" is limited. For example, both the infalling primordial gas and the outflowing metal-enriched gas could reduce the metallicity of the ISM. Without the extra constraints of the gas flow histories, models with different recipes for infall or outflow can predict similar present-day metallicities for a galaxy. Unfortunately, observations of gas inflow or outflow at any given time are difficult. Thus, the evolutionary history of the gas content is hard to constrain. In certain cases, extra assumptions are necessary to control CEM models based on the amount of gas or the history of gas entering. For example, when applying the bathtub model to the MaNGA galaxy, \citet{barrera18} assumed a quasi-equilibrium for the gas content so that the gas mass or fraction does not evolve significantly with time. Some models invoke a parameterized gas infall history, such as the infall models that have been successfully applied for a number of nearby disk galaxies \citep[e.g.,][]{yin09,chang12,kang12}.

Even though the evolutionary history of the gas content is not directly observable, the information on the SFH and MEH can be indirectly inferred from the spectra, as they are fossilized in the stellar population. Indeed, many studies have been devoted to deriving the precise SFHs of galaxies from their spectra using the stellar population synthesis (SPS) method \citep[e.g.][]{CidFernandes07,Wilkinson2015,Wilkinson17,sanchez16a,sanchez16b,sanchez22}. Although the SPS method is a powerful tool, it suffers from degeneracy among metallicity, age, and extinction and even varies with the algorithms adopted in the code \citep{sanchez11,walcher11}. In the series of work by Asa'd \& Goudfrooij \citep{asad20,asad21,asad22,goud21}, the authors discussed the limitation in depth and compared the results between the full-spectrum fitting and the synthetic color-magnitude diagram (CMD) method, which is based on comparing observed with theoretical CMDs created via Monte Carlo-based extractions from theoretical isochrone and is believed to provide relatively accurate MEH and SFH measurements. They found that, for simple stellar population-like star clusters, there is generally good agreement between the ages derived from CMDs and integrated spectra. However, for metallicity and reddening, the agreement between the results from analyses of CMDs and integrated spectra is significantly worse.

It is therefore very instructive to develop a CEM model that is regulated by the SFH instead of the gas infall history. In this paper, we explore such a model. In this model, we adopt the classic analytical chemical evolution framework but link all processes and the resulting parameters, including the MEH and gas content, to the SFH. This is different from many previous models, where the SFH is usually regulated by the gas content. The model is not physically motivated but may be more easily applicable to the observed data, as the SFH can be indirectly estimated from the optical spectral data even if it suffers degeneracy issues. The model also provides a physical link between the SFH and MEH. This can be very useful in SPS analysis, as it is a big nuisance that the SFH and MEH are degenerate if they are treated as independent parameters. 

The Local Cosmology from Isolated Dwarfs (LCID) project which has obtained deep Hubble Space Telescope (HST) observations of six isolated dwarf galaxies covering three types in the Local Group, provides us ideal targets for testing our model \citep{cole07,hidalgo09,hidalgo11,monelli10a,monelli10b,skm14}. Their CMDs have reached the oldest main-sequence turnoff points that make the precise measurements of SFH and MEH possible.
As they are isolated galaxies,  their evolution is not supposed to be affected by tidal stirring or ram pressure stripping, which are expected to remove gas and strongly affect their SFHs. Moreover, as dwarf galaxies, it is easier to reveal the role of SN feedback in the recycling of metals in CEMs.

The main content of the paper is as follows. In Section \ref{sec:CEM} we introduce our SFH-regulated model in detail; in Section \ref{sec:LCID}, we briefly introduce the LCID object and the observation constraints we used; in Section \ref{sec:results}, we present our results on LCIDs; in Section \ref{sec:dis}, we compare our SFH-regulated model with other models. We summarize in Section \ref{sec:sum}.

\section{The Chemical Evolution Model}
\label{sec:CEM}

\subsection{General CEM Framework}

The CEMs build a physical link between the metallicity of a galaxy and various gas cycling processes upon some hypothesis. By comparing model predictions with observational data, we may understand the creation and recycling of different elements in galaxies, which helps us to understand the evolutionary history of galaxies.

The CEM involves two levels of physical processes; one is the yield of metals from evolved stars, and the other is the mixing and recycling of metals within the ISM. Stars produce chemical elements through nuclear fusion reactions and then return these elements to the ISM when they die. According to the stellar evolution and nucleosynthesis theory, we know how much and which elements a star contributed and at what time, as described by the stellar yield \citep[e.g.][]{Woosley95,vandenHoek97,Iwamoto99} and the stellar mass lifetime relationship \citep[e.g.][]{larson74,rana91}.
Then, the metal abundance, which is defined as a metal-to-gas mass ratio, of galaxies is determined by both their SFH and the gas flow process. The SF process not only creates new stars but also traps metals from the ISM. Gas flow, which includes both inflow and outflow, can not only reduce the metal abundance but also expel metals from galaxies. Additionally, the star formation rate (SFR) of a galaxy is closely linked to its gas content, as indicated by the Kennicutt-Schmidt (KS) law \citep{ken98a,ken98b}.

To address different scientific questions, different CEMs adopt different assumptions or simplifications on these processes, and they can be solved numerically or analytically. The simplest CEM is to assume that there is no gas inflow and outflow \citep{searle72}. Therefore, the total mass of the galaxy does not change with time. This is the well-known closed-box (CB) model. The CB model describes a rapid collapse process, which is more suitable for early-type galaxies. Its metallicity depends only on the gas fraction, which is defined as the gas-to-total mass ratio. However, the CB model predicts a too-high metallicity at a given gas fraction \citep{garnett02, yin11}. It also predicts too many old metal-poor stars in the metallicity distribution function of G-dwarf stars, which contradicts the observation in the solar neighborhood (the so-called ``G-dwarf problem") \citep{bergh62, schmidt63, audouze76}. To slow down the enrichment rate and solve the G-dwarf Problem, gas inflow and/or outflow should be taken into account. The leaky box (LKB) model \citep{zhu17} provokes a metal-enriched outflow to reduce the metallicity. On the other hand, the gas infall model \citep{yin09,kang12} considers a scenario in which a galaxy continuously accretes the primordial gas and gradually builds up its stars. The IF model can dramatically reduce the number of metal-poor stars formed at early times, thus alleviating the G-dwarf problem \citep{yin09}. The bathtub model includes both the gas infall and outflow processes and therefore provides a more general framework \citep[e.g.][]{dave12,lilly13,barrera18}. 

The basic equations that describe the CEM framework are as follows.

(1) The time evolution of the surface density of the total baryonic mass $\Sigma_{\text{tot}}(t)$ (including the stars and gas) is regulated by the infall and outflow of the gas as
\begin{equation}\label{eq:GCE-tot}
    \dot\Sigma_{\text{tot}}(t) \equiv  \dot\Sigma_{*}(t)+\dot\Sigma_{\text{gas}}(t) 
                     = f_{\text{in}}(t)-f_{\text{out}}(t)  .   
\end{equation} 

(2) The time evolution of the gaseous surface density $\Sigma_{\text{gas}}(t)$ is given by
\begin{eqnarray}\label{eq:GCE-gas}
    \dot\Sigma_{\text{gas}}(t) & = & \dot\Sigma_{\text{tot}}(t) - \dot\Sigma_{*}(t) \nonumber  \\
                 & = &  f_{\text{in}}(t)-f_{\text{out}}(t) - \Psi(t) + E(t) ,
\end{eqnarray}
where $\Psi(t)$ is the surface density of SFR, and $E(t)$ is the surface density of the mass return rate due to the dying stars ejecting both the enriched and unenriched material back into the ISM. Assuming that stars more massive than 1$M_{\odot}$ die instantaneously and return gas to the gas reservoir (instantaneous recycling approximation, IRA), $E(t)$ can be simplified as
\begin{equation}
    E(t)=R~\Psi(t)
\end{equation} 
where $R$ is the mass return fraction, and $R=0.3$ if a Salpeter initial mass function (IMF)\citep{salp55} is assumed.

(3) If the enriched gas is ejected and well mixed with the surrounding ISM instantaneously (instantaneous mixing approximation, IMA), the time evolution of the mass of the element $i$ in the gas is described by the equation
\begin{eqnarray} \label{eq:GCE-Z0}
    \frac{d[Z_i(t)\Sigma_{\text{gas}}(t)]}{dt} & = & -Z_i(t)\Psi(t) + E_i(t) \nonumber \\
                            &  & + Z_{\text{in},i}f_{\text{in}}(t) -Z_{\text{out},i}f_{\text{out}}(t)
\end{eqnarray} 
where $Z_i$ is the mass fraction of element $i$ in the gas, that is, metallicity by mass, and $Z_{\text{in},i}$ and $Z_{\text{out},i}$ represent the metallicity of element $i$ in the infall and outflow gas, respectively. 
In this equation, the first term is the metal locked by the newly formed stellar generation. The second term, $E_i$, is the corresponding newly produced metals that are ejected back into the ISM. The last two terms are the metals brought in by the infalling gas and blown out by the outflowing gas, respectively. The metallicity of the infalling gas $Z_{\text{in},i}$ is set to zero when the gas is assumed to be primordial, and the metallicity of the outflow gas $Z_{\text{out},i}$ is set to $Z_i$ when IMA is assumed.

The return mass $E_i$ integrates the yields $y_i$ of elements $i$ over the stellar mass according to its forming rate. The yield $y_i$ depends not only on the element $i$ but also on the stellar mass. Low- and intermediate-mass stars ($0.8-8 M_{\odot}$) produce He, N, C, and heavy s-process elements and die as C-O white dwarfs or Type Ia SNe, depending on whether they are single or binary stars. Massive stars ($>8-10 M_{\odot}$) produce mainly $\alpha$-elements (O, Ne, Mg, S, Si, Ca), some Fe, light s-process elements and perhaps r-process elements and explode as core-collapse SNe \citep{matteucci16}.

If instantaneous cycling and mixing assumptions are adopted, Equation~(\ref{eq:GCE-Z0}) can be simplified as
\begin{eqnarray} \label{eq:GCE-Z}
    \frac{d[Z(t)\Sigma_{\text{gas}}(t)]}{dt} & = & [y-Z(t)](1-R)\Psi(t)  \nonumber\\
                            &  & + Z_{\text{in}}f_{\text{in}}(t) -Z_{\text{out}}f_{\text{out}}(t)
\end{eqnarray} 
where $Z$ is the global metallicity, and $y$ is the total yield per stellar generation, which can be obtained by integrating the stellar yield with the IMF that describes the distribution of stars at birth as a function of the stellar mass.

With IRA assumption, the CEM equation can be solved analytically or numerically. However, it is a poor approximation for chemical elements produced on long timescales such as C, N, and Fe. For elements like O, which are almost entirely produced by short-lived core-collapse SNe, IRA is an acceptable approximation. 

\subsection{Our SFH-regulated model}
\label{sec:CEM_our}

The model we propose in this work follows the same framework as the bathtub model \citep{dave12,lilly13,barrera18}; it has a gas reservoir and considers both the infall and outflow processes. However, we directly relate the gas assembly history $\dot\Sigma_{\text{gas}}(t)$, the inflow rate $f_{\text{in}}(t)$, and the outflow rate $f_{\text{out}}(t)$ to the SFH of the galaxy. The model constructed in this way is to some extent more straightforward, as SFH can be more easily constrained from observations (e.g., through the CMD or spectrum fitting). We call our model ``the SFH-regulated model", and explain it in detail below.

In this model, we derive the time variation of all variables. Most importantly, we infer the MEH of a certain galaxy from its measured SFH (upper panel of Figure 1), and then compare this predicted MEH with the measured one (lower panel of Figure 1). 

In this study, we simply take the global metallicity [M/H], which is defined as $\log(Z/Z_{\odot})$, where $Z_{\odot} = 0.02$ is usually assumed, and trace its evolution with the gas recycling process of galaxies. 
The reason is as follows.
On the one hand, in observations, especially in the field of extragalactic astronomy, it is common to obtain the global metallicity of a galaxy rather than the individual elements. For example, in the standard and popular full-spectrum fitting of galaxies that aims to constrain the stellar population (SFH and MEH) of galaxies, the derived MEH is the global metallicity of each theoretical simple stellar population. However, this metallicity is affected by the degeneracy with the age of stellar populations. 
On the other hand,  for the observational constraints in this work (see Sec. \ref{sec:LCID_SFH}), which are based on a comparison of observed CMDs with theoretical ones created via Monte Carlo-based extractions from theoretical isochrones, the metallicities of the isochrone templates represent the global metallicity too. 

Therefore, in our simple CEM, we do not distinguish between individual elements but instead use the total yield, which is scaled to twice the yield of oxygen $y_O$, to trace the evolution of global metallicity, where instant recycling and mixing assumptions are also adopted.

\subsubsection{The SF Law and Gas Assembling History}

We adopt the KS law \citep{ken98a,ken98b}  with a KS index $n \sim 1.4$ and assume that the SF follows  this law in the whole lifetime of the galaxy,
\begin{equation}\label{eq:kslaw}
    \Psi=\epsilon\Sigma_{\text{gas}}^n~,
\end{equation}
where $\epsilon=0.25$ is the efficiency.

Under this assumption, the evolution history of the gas surface density can be estimated through
\begin{equation}\label{eq:sflaw}
    \Sigma_{\text{gas}}(t)= \left[\frac{\Psi(t)}{\epsilon}\right] ^{1/n}~,
\end{equation}
where $\Sigma_{\text{gas}}$ and $\Psi$ are in units of \mspc~and \mspcgyr, respectively.

\subsubsection{The Net Gas Flow}
\label{sec:CEM_our_fnet}

How the gas mass surface density changes over time depends on the gas inflow/outflow and SF processes, as described by Equation (3). 
We define a net gas flow rate $f_{\text{net}}(t)$, which describes the competition results of the gas inflow and outflow as
\begin{equation}\label{eq:fnet2}
    f_{\text{net}}(t) \equiv f_{\text{in}}(t) - f_{\text{out}}(t).
\end{equation}

Combined with Equation (3), we notice that for any given SFH, a net gas flow rate $f_{\text{net}}(t)$ can be estimated as
\begin{equation}\label{eq:fnet}
  \begin{split}
    f_{\text{net}}(t) =\frac{1}{n\cdot\epsilon^{1/n}}\cdot \Psi(t)^{\frac{1}{n}-1} \cdot\dot\Psi(t) +(1-R)\Psi(t).
  \end{split}
\end{equation}
For a given SFR, $f_{\text{net}}(t)>0$ means that a net gas accretion from the halo to its galaxy is needed, whereas $f_{\text{net}}(t)<0$ implies that there is a net gas outflow from the galaxy to its halo. In other words, for a given SFH, the net flow rate describes the {\it minimum} gas exchange rate required.

\subsubsection{The Outflow Rate}

\citet{guo11,guo13} proposed that part of the energy released by the SN explosions reheats the surrounding gas from the ISM to a hot gas halo, and the efficiency is halo-dependent,
\begin{equation}\label{eq:fout}
    \epsilon \propto \left[0.5+\left(\frac{v_{\text{vir}}}{v_{\text{norm}}}\right)^{-\beta}\right],
\end{equation}
where $v_{\text{norm}} = 70$  km/s and $\beta = 3.5$ are proposed in their preferred model.
In our model, we set our SN-driven outflow rate $f_{\text{wd}}$ to be proportional to the SFR and relate it to the gravitational potential of the host galaxy following \citet{guo11,guo13}, 
\begin{equation}\label{eq:fwd}
    f_{\text{wd}}(t)=\eta \left[0.5+\left(\frac{v_{\text{vir}}}{70 ~km~s^{-1}}\right)^{-3.5}\right]\cdot\Psi(t)~,
\end{equation}
where $v_{\text{vir}}$ is the virial velocity (Section \ref{sec:LCID_vvir}), and $\eta$ is the wind efficiency. We note that $\eta$ is the only free parameter in our model. Equation~(\ref{eq:fwd}) considers the effect of the host galaxies' potential in a way that the lower-mass galaxy tends to have stronger outflow for a given SFR. 

Sometimes the $f_{\text{wd}}$ derived by Equation~(\ref{eq:fwd}) is weaker than the net outflow rate $f_{\text{net}}$ required by Equation (\ref{eq:fnet})(it usually relates to a very sharp shutdown of SF). This implies that there are additional factors that affect the outflow process, such as the environment, active galactic nucleus, etc. In this case, we force the outflow rate of the whole system $f_{\text{out}}$ to be equal to the absolute net outflow rate, 
\begin{equation}\label{eq:fout2}
    f_{\text{out}}(t)=\left\{ \begin{array}{ll}
       f_{\text{wd}}(t)  &  if~|f_{\text{net}}| < f_{\text{wd}} \\
       |f_{\text{net}}(t)|  & if~|f_{\text{net}}| > f_{\text{wd}}
    \end{array} \right.~.
\end{equation}

\subsubsection{The inflow rate}

In the real case, the accretion of the cooling gas and the ejection of the metal-enriched gas could happen simultaneously. The net flow rate is the result of the balance between the infall and outflow processes. Combining Eqs. (\ref{eq:fnet2})-(\ref{eq:fout2}), the inflow rate is estimated as
\begin{equation}\label{eq:finlcid}
  \begin{split}
    f_{\text{in}}(t) &= f_{\text{net}}(t) + f_{\text{out}}(t) \\
    &=\frac{1}{n\cdot\epsilon^{1/n}}\cdot \Psi(t)^{\frac{1}{n}-1} \cdot\frac{d\Psi(t)}{dt} +(1-R)\Psi(t) \\
              &+ \eta \left[0.5+\left(\frac{v_{\text{vir}}}{70 ~km~s^{-1}}\right)^{-3.5}\right]\Psi(t) ~.
  \end{split}
\end{equation}
As shown by Equation (\ref{eq:finlcid}), once the SFH is given, a strong outflow rate (large $\eta$) implies that the system requires a large amount of gas falling into the system to compensate for its loss and keep the SF.

\subsubsection{The Metal Enrichment History}
\label{sec:CEM_MEH}

Based on the scenario we described above, we can calculate the MEH $Z(t)$ of any given SFH $\Psi(t)$.

We assume that the infalling gas is primordial, i.e., $Z_{\text{in}}=0$, and the outflow gas has the same metal component as the ISM at that time, that is, $Z_{\text{out}}=Z(t)$. Based on these assumptions, the evolution of the global metallicity in the gaseous phase is described by
\begin{equation}\label{eq:z}
  \begin{split}
    &\frac{d(Z\Sigma_{\text{gas}})}{dt} =-Z(1-R)\Psi(t)+y(1-R)\Psi(t)-Zf_{\text{out}} \\
    &=\left\{(y-Z)(1-R)-Z\eta\left[0.5+\left(\frac{v_{\text{vir}}}{70 ~km~s^{-1}}\right)^{-3.5}\right]\right\}\Psi(t)\,,
  \end{split}
\end{equation}
where $y$ is the yield. In this work, we adopt $y=0.02$ to calculate the global metallicity of the galaxies.

So far, all variables ($\Sigma_{\text{gas}}$, $\Sigma_*$, $f_{\text{in}}$, $f_{\text{out}}$, $Z$) are directly related to the SFH and expressed as a function of $t$. Once the SFH is known, $\eta$ is the only free parameter in the model. That is to say, in our CEM, the MEH is physically determined by the SFH. In the next section, we validate this SFH-regulated CEM with the observational data.

\section{LCID galaxies}
\label{sec:LCID}

Taking advantage of the great sensitivity and spatial resolution of the Advanced Camera for Surveys  (ACS) on board the HST, the LCID project obtained the first CMDs reaching the oldest main-sequence turnoff stars with good photometric accuracy for six relatively isolated dwarf galaxies in the Local Group that could provide an accurate SFH going back to the earliest times.
The LCID galaxies include two dwarf irregular (dIrr) galaxies, IC 1613 and Leo A \citep{cole07,skm14}; two dwarf transition-type (dTran) galaxies, LGS 3 and Phoenix \citep{hidalgo09,hidalgo11}; and two dwarf spheroidal (dSph) galaxies, Cetus and Tucana \citep{monelli10a,monelli10b}. 

\begin{deluxetable*}{llcccccccc}
 \tablewidth{0pt}
 %\tabletypesize{\small}
 \tablecaption{Observational properties of LCID galaxies. \label{Tab:lcid}}
 \tablehead{
  \colhead{Name} &
  \colhead{Type} &
  \colhead{$M_{*}^{a,b}$} &
  \colhead{$r_{50}^b$} &
  \colhead{$\Sigma_{*,50}$} &
  \colhead{$M_\text{\hi}^c$} &
  \colhead{$\mu$} &
  \colhead{$r_{\text{vir}}$} &
  \colhead{$v_{\text{vir}}$} &
  \colhead{[M/H]}  \\
  \colhead{ } &
  \colhead{ } &
  \colhead{($10^6$\ms)} &
  \colhead{(pc)} &
  \colhead{(\mspc)} &
  \colhead{($10^6$\ms)} &
  \colhead{} &
  \colhead{(kpc)} &
  \colhead{(km/s)} &
  \colhead{ }  \\
  \colhead{} &
  \colhead{(1)} &
  \colhead{(2)} &
  \colhead{(3)} &
  \colhead{(4)} &
  \colhead{(5)} &
  \colhead{(6)} &
  \colhead{(7)} &
  \colhead{(8)} &
  \colhead{(9)} }
 \startdata
  Cetus  & dSph  & 7.0       & 703        & 0.88             & $<0.29^d$  & $<$0.04 & 54.3      & 30.2      & -1.6 $\sim$ -1.5 $^e$ \\
  Tucana  & dSph  & 3.2       & 284        & 1.11            & $<0.39^d$  & $<$0.11 & 46.0      & 27.8      & -1.95 $\sim$ -1.74 $^f$ \\
  LGS 3   & dTran & 1.9       & 470        & 0.69            & 0.38       & 0.21  & 48.8      & 26.3      & -2.3 $\sim$ -2.0 $^e$ \\
  Phoenix & dTran & 3.2       & 454        & 0.59            & 0.12       & 0.05  & 47.6      & 27.8      & -1.17$^g$ \\
  IC~1613 & dIrr  & 100.0     & 1496       & 7.11            & 65.0       & 0.46  & 80.4      & 40.3      & -1.3$^h$ \\
  Leo~A   & dIrr  & 6.0       & 499        & 3.84            & 11.0       & 0.71  & 59.4      & 29.7      &  -1.4$^i$ \\
\enddata
\tablenotetext{}{Notes. (1) Galaxy type, (2) total stellar mass, (3) half-light radius, (4) averaged stellar surface density within $r_{50}$ calculated through Equation (\ref{eq:s50}), (5) the total \hi gas mass, (6) total gas fraction calculated by $\mu=\frac{1.33M_\text{\hi}}{1.33M_\text{\hi}+M_*}$, (7) viral radius calculated through Equation (\ref{eq:rvir}), (8) the viral velocity estimated in this work calculated through Equation (\ref{eq:vvir});  (9) mean global metallicity [M/H], defined as $\log(Z/Z_{\odot})$. \\
$^a$\citet{gallart15}. \\
$^b$\citet{mccon12}. \\
$^c$\citet{weisz14}. \\
$^d$\citet{grcevich16}. \\
$^e$\citet{mccon05}. \\
$^f$It is calibrated from [Fe/H]\citep{Fraternali09} according to Equation(1) of \citet{Ferraro99} and assuming [$\alpha$/Fe] ranges from zero to 0.3.\\
$^g$\citet{Ross15}.\\
$^h$\citet{skm14}.\\
$^i$\citet{cole07}.}
\end{deluxetable*}

The main global properties of these six LCID galaxies are listed in Table \ref{Tab:lcid},%\ref{Tab:lcid},
including the galaxy type (column (1)), the total stellar mass $M_*$ (column (2)), the half-light radius $r_{50}$ from \citet{mccon12} (column (3)), the averaged stellar surface density $\Sigma_{*,50}$ within $r_{50}$ (column (4)), the total \hi mass $M_\text{H I}$ (column (5); for both Cetus and Tucana, which have no \hi detected, the detection limits of HIPASS at their distances are listed; \citet{grcevich16}), the total gas fraction calculated from the total \hi mass and stellar mass (column (6)), the virial radius $r_{\text{vir}}$ (column (7)) and virial velocity $v_{\text{vir}}$ estimated in this work (column (8); see Section \ref{sec:LCID_vvir} for details), and the mean global metallicity [M/H] (column (9)).

\subsection{SFHs and MEHs}
\label{sec:LCID_SFH}

The quantitative SFH of a composite stellar population can be derived from a CMD by comparing the observed CMD with the synthetic one \citep[e.g.][]{tolstoy09}. The model predictions are derived from theoretical isochrones combining a prescription for the IMF, the influence of unresolved binary stars, observational errors, etc. The variation of SFR with time, i.e. the SFH, and the evolution of stellar metallicity, i.e. the MEH, are taken from the combination of parameters that results in the best reproduction of the observed CMD.

\begin{figure}[!t]
  \centering
  \includegraphics[width=9cm]{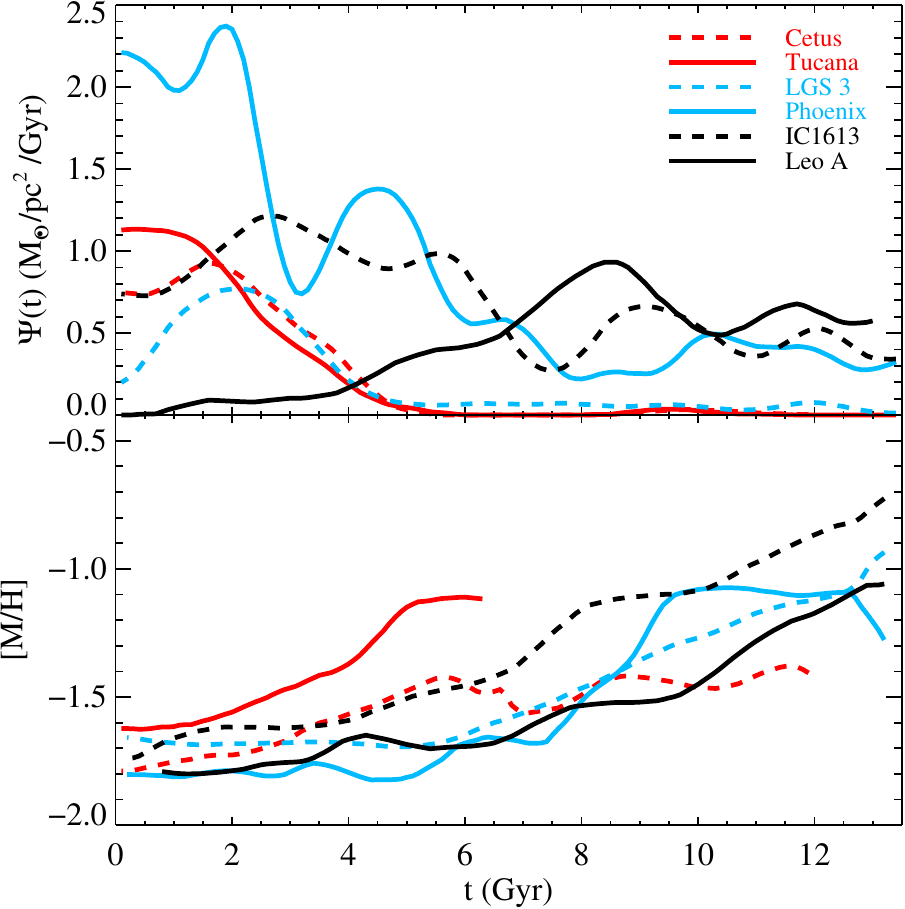}\\
    \caption{Observed SFHs and MEHs of six LCID galaxies. Two dSphs, Cetus and Tucana, are shown with red lines \citep{monelli10a,monelli10b}; two dTrans, LGS 3 and Phoenix, are shown with blue lines \citep{hidalgo09,hidalgo11}; and two dIrrs, IC 1613 and Leo A, are shown with black lines \citep{cole07,skm14}. The upper panel shows the evolution histories of SFR surface density, whereas the lower panel shows the MEHs.}
  \label{Fig:SFHobs}
\end{figure}

The CMD synthesis method gives a reliable three-dimensional SFH, $\psi(t, Z)$, for each galaxy. The two projections of $\psi(t, Z)$ are the age distribution of $\psi(t, Z)$ integrated over metallicity and the metallicity distribution of $\psi(t, Z)$ integrated over SFR, i.e. the SFH and MEH in our model (see Figure \ref{Fig:SFHobs}). For more details on the SFH and MEH determinations of these galaxies, refer to the above-referenced papers.

The evolutions of the SFR surface densities of these LCID galaxies have been illustrated in a series of papers \citep{hidalgo09, hidalgo11, monelli10a, monelli10b, skm14}. We show their SFHs in Figure \ref{Fig:SFHobs}. The two dSphs formed most of their stars at quite an early epoch. On the contrary, the dTrans and dIrrs continued forming stars until the present day, although the dTrans are at a lower rate. 

The SFH $\Psi(t)$ in the above papers are all measurements of the surface density (in units of \mspcgyr), except for Leo A whose total SFR (in units of \msyr) was measured by \citet{cole07}. Therefore, to be consistent with all six galaxies, we estimate the time evolution history of the average SFR surface density for Leo A below. 

We assume that the average SFR surface density $\Psi(t)$ within $r_{50}$ of Leo A follows the same time evolution trend as the total SFR $\psi_{\text{Cole07}}(t)$, 
\begin{equation}\label{eq:Leoa1}
  \begin{split}
    \Psi^{\text{LeoA}}(t)=A\cdot \psi^{\text{Cole07}}(t)~,
  \end{split}
\end{equation}
where $A$ is the scale parameter. The integrated $\Psi^{\text{LeoA}}(t)$ over the cosmic time is constrained by the stars alive today,
\begin{equation}\label{eq:Leoa2}
  \begin{split}
    (1-R)\int \Psi(t)^{\text{LeoA}} dt=\Sigma_{*,50}~,
  \end{split}
\end{equation}
where $\Sigma_{*,50}$ is the average stellar surface density within the half-light radius in Leo A and can be estimated from the total stellar mass $M_{*}$ and the half-light radius $r_{50}$ from \citet{mccon12},
\begin{equation}\label{eq:s50}
  \begin{split}
    \Sigma_{*,50}=\frac{0.5 M_{*}}{\pi r_{50}^2} ~.
  \end{split}
\end{equation}
Therefore, combining Eqs.~(\ref{eq:Leoa1})-(\ref{eq:s50}), we can calculate the scale parameter $A$ and then derive the evolution of the averaged SFR surface density of Leo A as
\begin{equation}\label{eq:Leoa3}
  \begin{split}
    \Psi^{\text{LeoA}}(t) = \frac{\Sigma_{*,50}}{(1-R)\int \psi^{\text{Cole07}}(t)dt}\cdot \psi^{\text{Cole07}}(t)
  \end{split}
\end{equation}
where $R$ is the mass return fraction (i.e., the fraction of mass that is ejected into the ISM when stars die). In our study, we assume $R=0.3$ based on the IRA assumption and the Salpeter IMF \citep{salp55}.

The time evolutions of the SFR surface densities of all six LCID dwarf galaxies are plotted, including the estimated one for Leo A, in the upper panel of Figure  \ref{Fig:SFHobs}. 

The MEH derived by the LCID group was obtained from the averaged metallicity of the three-dimensional SFH $\psi(t, Z)$ for each age. In the lower panel of Figure  \ref{Fig:SFHobs}, we show all the MEHs for the six galaxies.

However, considering the old luminous blue stragglers observed in Tucana could be interpreted as a small later burst of SF at $t\sim9.5$ Gyr by the CMD fitting code, \citet{monelli10a} claimed that the feature of the MEH based on these stellar populations are unreal and did not show the derived MEH of Tucana after $t=6$ Gyr in their Figure 6.

\subsection{The Virial Velocity}
\label{sec:LCID_vvir}

\citet{brook14} provided the relation between the stellar mass of Local Group galaxies and the halo masses in which they are hosted assuming a $\Lambda$CDM cosmology,
\begin{equation}
    M_{\text{halo}}=M_{*}^{0.32}\times M_0 ~,
\end{equation}
where $M_0=7.96\times10^7$\ms. The virial radius can be estimated from the halo mass as
\begin{equation}\label{eq:rvir}
    r_{\text{vir}}=\left(\frac{M_{\text{halo}}}{\frac{4}{3}\pi\rho_{\text{vir}}}\right)^{\frac{1}{3}} ~~~\text{Mpc}~,
\end{equation}
where $\rho_{\text{vir}}=2.84\times10^{13}~(M_{\odot}/h)/(Mpc/h)^3$ with $h=0.7$ (Plank 2015 cosmology).

Once we have both the halo mass $M_{\text{halo}}$ and the virial radius $r_{\text{vir}}$, the virial velocity $v_{\text{vir}}$ can be estimated to be
\begin{equation}\label{eq:vvir}
    v_{\text{vir}}=\sqrt{\frac{G M_{\text{halo}}}{r_{\text{vir}}}}~.
\end{equation}
We list the virial radius $r_{\text{vir}}$ and the estimated virial velocity $v_{\text{vir}}$ (via Equation(\ref{eq:vvir})) for six LCID galaxies in Table \ref{Tab:lcid}.

\section{Our Model Fitting to the LCID Galaxies}
\label{sec:results}

Based on the CEM we described in Section \ref{sec:CEM}, for any given $\Psi(t)$, we can predict the time evolution of several main properties of the galaxy, such as the evolution histories of the gas mass surface density $\Sigma_{\text{gas}}(t)$, the gas inflow rate $f_{\text{in}}(t)$, the outflow rate $f_{\text{out}}(t)$, and the MEH $Z(t)$. 

For LCID galaxies, we use $\Psi(t)$ determined from the CMD synthesis method (upper panel of Figure~\ref{Fig:SFHobs}) to derive their MEHs $Z(t)$ and then compare with the measured ones (lower panel of Figure~\ref{Fig:SFHobs}).
Below, we show the detailed predicted evolutionary histories for all 6 LCID galaxies.

\subsection{Early-type Dwarf Galaxies}
\label{sec:results_dSph}

\begin{figure*}[!t]
  \centering
  \includegraphics[width=15cm]{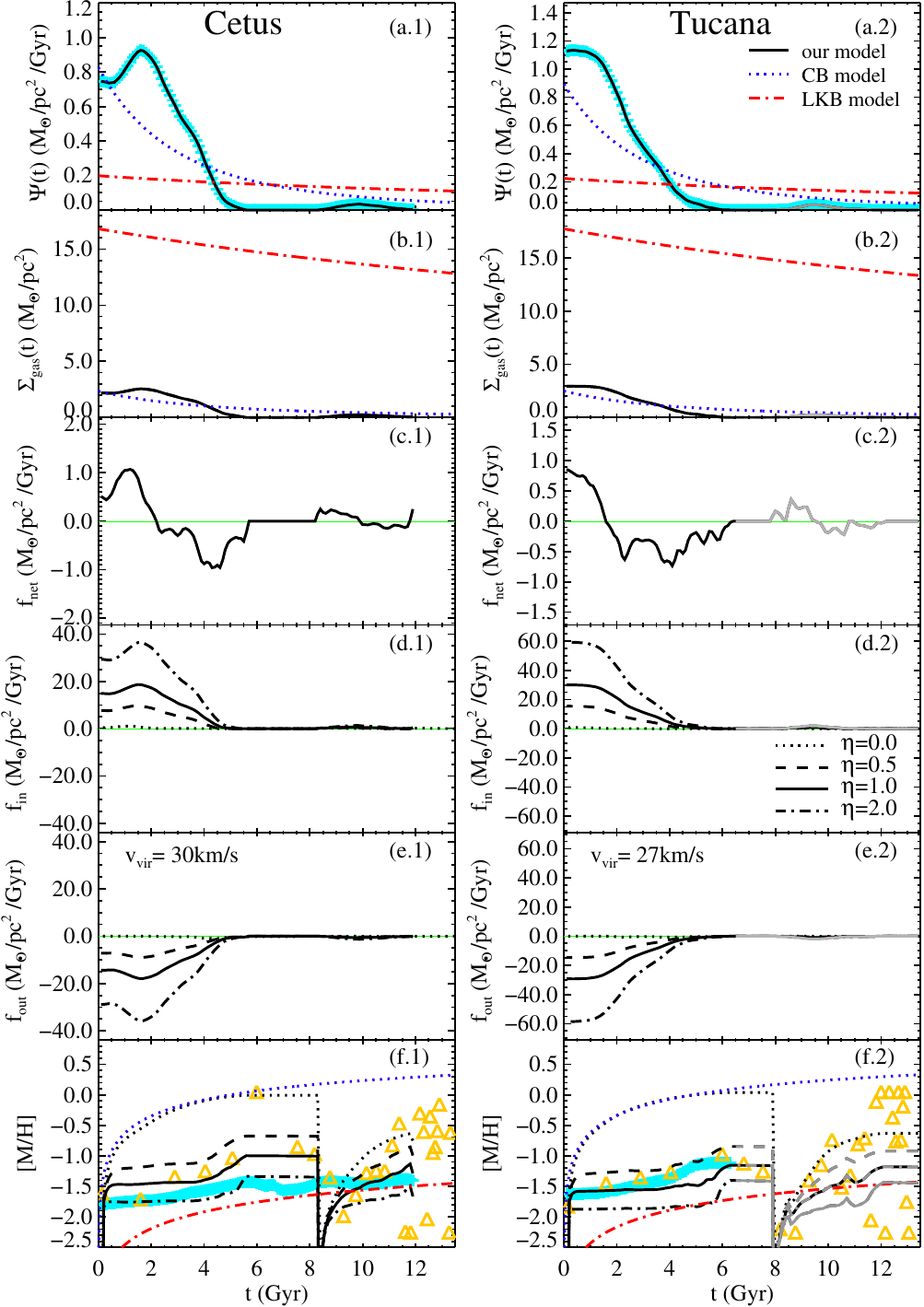}\\
    \caption{Time evolution of the dSph galaxies Cetus (left column) and Tucana (right column). The SFHs measured by the CMD fitting method are shown in cyan \citep{skm14} in panels (a.1) and (a.2), and the black solid lines are used as the input SFHs in our model. The time evolution of gas surface density is shown in panels (b.1) and (b.2). In panels (c.1) and (c.2), the black solid lines represent the net gas flow rate. The gas inflow and outflow histories are shown in panels (d) and (e), and the MEHs predicted by our model (black lines) compared with those measured by \cite{skm14} (cyan) and Dan Weisz (private communication; yellow triangles) are shown in panel (f). The black dotted, dashed, solid, and dashed-dotted lines indicate the $\eta=0.0, 0.5, 1.0,$ and $2.0$ cases, respectively. In the case of Tucana, all of the evolution histories are shown as gray after $t \sim 6$ Gyr. We also overplotted the evolution histories predicted by the CB (blue dotted lines) and LKB model (red dashed-dotted lines) models as a comparison (see Section \ref{sec:otherCEM} for more details).}
  \label{Fig:dsph}
\end{figure*}

\begin{figure*}[!tbh]
  \centering
  \includegraphics[width=16cm]{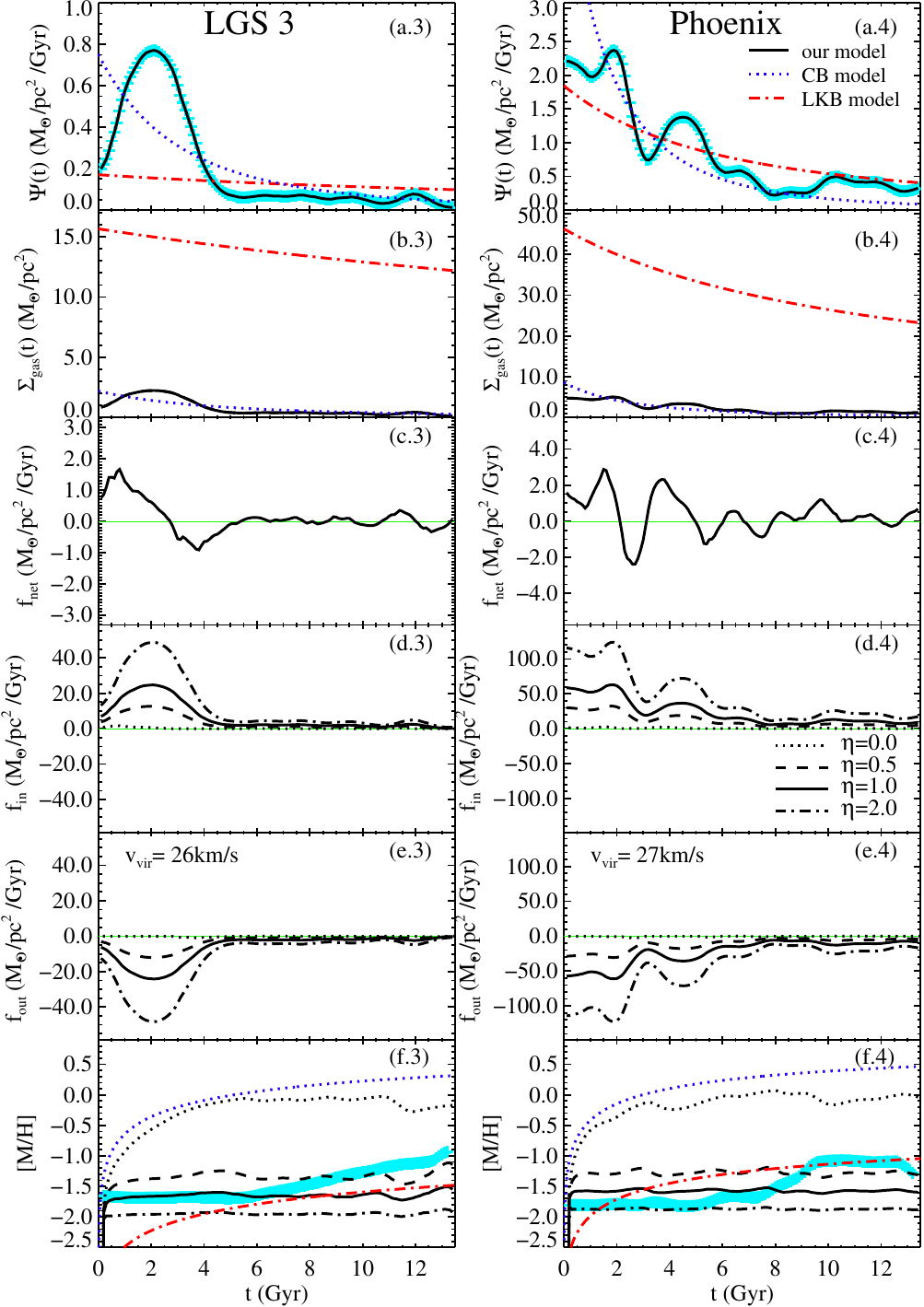}\\
    \caption{Time evolution of dTran galaxies LGS 3 and Phoenix. Similar to Figure  \ref{Fig:dsph}, this figure shows the evolution of the SFR, the gas surface density, the net gas flow rate, the gas inflow and outflow rates, and the MEH from top to bottom.}
  \label{Fig:dtran}
\end{figure*}

\begin{figure*}[!tbh]
  \centering
  \includegraphics[width=16cm]{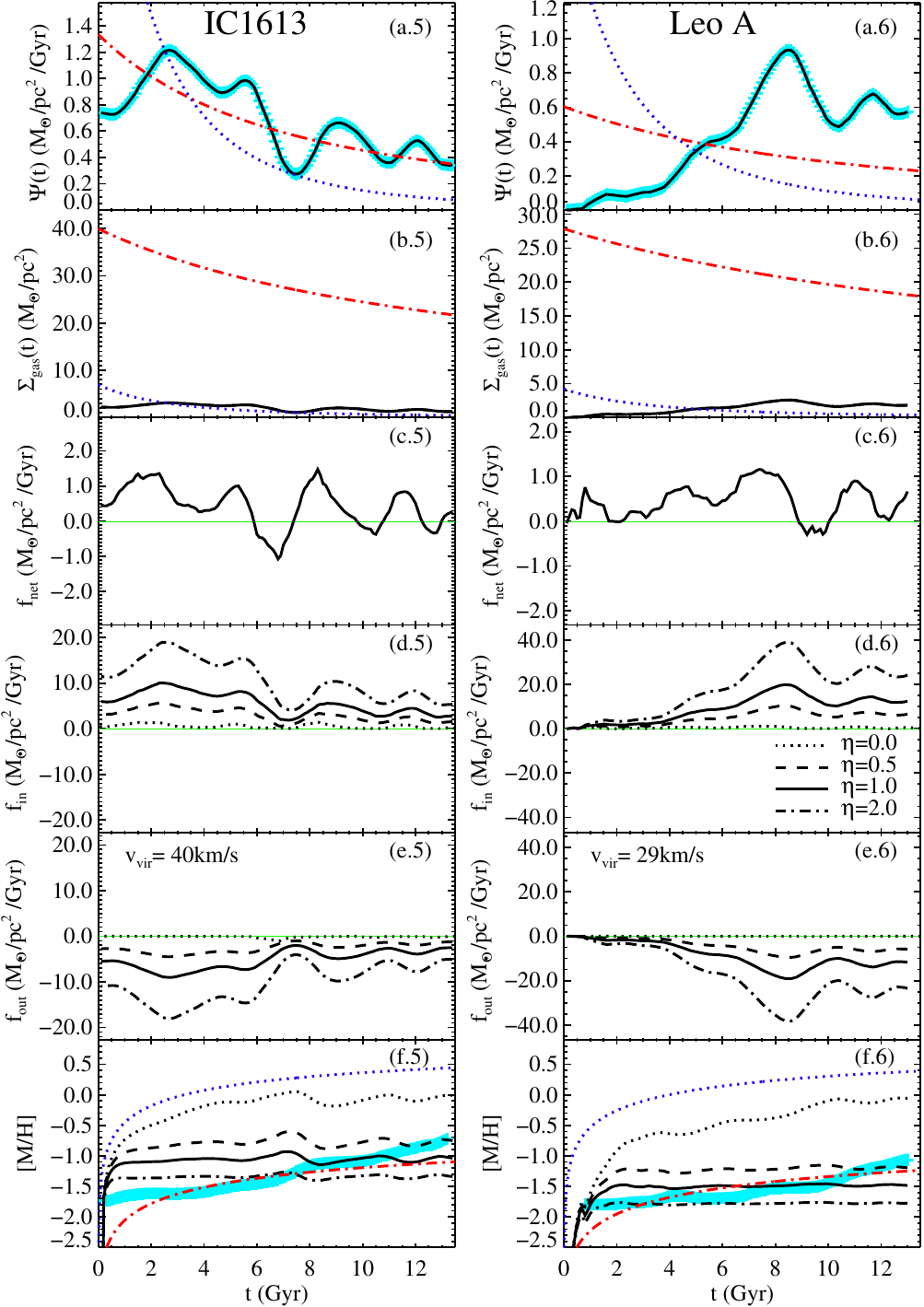}\\
    \caption{Time evolution of dIrr galaxies IC 1613 and Leo A. Similar to Figure  \ref{Fig:dsph}, this figure shows the evolution of the SFR, the gas surface density, the net gas flow rate, the gas inflow and outflow rates, and the MEH from top to bottom.}
  \label{Fig:dIrr}
\end{figure*}

Early-type dwarfs are those that started their evolution with a dominant but short ($<$few Gyr) SF event at an early stage. Cetus and Tucana are two predominantly old and metal-poor dSph galaxies. The bulk of the SF occurred at old ages, and most of the stars were formed in the first 4 Gyrs with a peak at around $t\sim 1$ Gyrs \citep{monelli10a}. Then the SF declined and stopped around $t\sim4-5$ Gyr. Later on, there is a small peak of SF predicted at $t\simeq 9.5$ Gyr. The galaxies are quenched and turn into red.

In Figure ~\ref{Fig:dsph}, we show our results from modeling Cetus (left column) and Tucana (right column). 
Using the measured SFH as input (solid line in panel (a) of Figure~\ref{Fig:dsph}), we estimate how the gas surface density $\Sigma_{\text{gas}}$ changes with time according to Equation(\ref{eq:sflaw}) and show the model prediction in panel (b) of Figure ~\ref{Fig:dsph}. The results show that the variation trend of the gas surface density is similar to that of the SFR. After the dSphs have consumed all of their gas that was initially gathered, they stay in a gas-poor quiescent state that lasts until a second peak of the gas surface density shows up around $t \sim 9.5$ Gyr. In our model, this second peak implies that, in order to produce the SF kick at $t \sim 9.5$ Gyr, there should be some fresh gas cooling down and fueling the galaxy at that time.

The net flow rate is the result of the competition between the inflow and outflow; therefore, it shows the minimum inflow/outflow rate required (see Section \ref{sec:CEM_our_fnet}). With the evolution curves of $\Sigma_{\text{gas}}(t)$ and $\Psi(t)$, we derive the net gas flow rate $f_{\text{net}}$ through Equation (\ref{eq:fnet}) and show it with black solid lines in panel (c) of Figure ~\ref{Fig:dsph}.  We note that a positive value of $f_{\text{net}}$ implies a net inflow, whereas a negative value implies a net outflow. As early-type dwarf galaxies, both Cetus and Tucana were dominated by gas accretion in the very first few gigayears (net inflow); then, the gas cycling process is dominated by the gas outflow until the SF is quenched (net outflow).

If assuming no extra stellar wind, i.e., $\eta=0.0$ in Equation (\ref{eq:fwd}), the gas inflow rate $f_{\text{in}}$ is equal to the positive part of $f_{\text{net}}$, and the gas outflow rate $f_{\text{out}}$ is the negative part of $f_{\text{net}}$. The gas was accreted into the galaxy in the first $\sim 2$ Gyr, then started to flow out during the next 4 Gyr. After an $\sim 2$ Gyr quiescent period, the second gas accretion process occurred. Note that even in the $\eta = 0.0$ case, gas outflow can happen in a galaxy. The $\eta = 0.0$ case only represents a scenario in which there is no additional SFR-related wind in the model. When $f_{\text{net}} < 0$, the outflow has to happen in our model. 

The high SFR measured at the early time may have injected enough energy into the ISM to blow out a significant fraction of the initial gas content. We set a nonzero $\eta$ to include the effect of the SFR-driven outflow in Figure ~\ref{Fig:dsph}. The outflow is strong at the early time because of the high SFR at that time. It decreases gradually afterward (panel (e) of Figure ~\ref{Fig:dsph}) due to the decrease of the SFR. To compensate for such a mass loss, our model requires a time-decreasing inflow, which is shown in panel (d). After this first gas cycling episode, the galaxies turn into a quiet phase from $t \sim 6$ to 8 Gyr. After $t \sim 8$ Gyr, our model predicts another gas accretion event, which induces the second SF and outflow episode.  

The gas cycling process determines the MEH. The gas-phase metallicity can be inferred from Equation (\ref{eq:z}), and we show the results in panes (f) of Figure~\ref{Fig:dsph}. The measured MEH is represented by the colored symbols; cyan is the result of \cite{skm14}, and the yellow triangles are from D. Weisz (2023, private communication). The model predictions are plotted as lines. Our model predicts that the metallicity [M/H] reaches $\sim -1.7$ dex very quickly, then slows down (panel (f) of Figure~\ref{Fig:dsph}). This trend is very consistent with the observation. It is clear that the gas-phase metallicity is sensitive to the outflow efficiency $\eta$. When the wind efficiency $\eta$ is set to 0.0 (black dotted lines in panel (f)), the metallicity increases very quickly, to much higher than the observed value. When the value of $\eta$ increases, the enrichment rate is highly depressed. Our predictions in panel (f) are comparable to the observations when $\eta\sim 1.0$, represented by the black solid lines. This implies that the outflow of enriched gas combined with the inflow of primordial gas does play a key role. The outflow decreases the effective yield of the stars, as suggested by \citet{garnett02}.

Another notable feature of panel (f) is that the predicted metallicity drops quickly at $\sim$ 8 Gyr. This is caused by the predicted gas accretion event that happened since $t \sim$ 8 Gyr (panel (d)), which is associated with the small SF bump in the input SFH (panel (a)). The dSphs have exhausted all the gas in their first SF episode ($t < 6$ Gyr) and stay in an extremely gas-poor state. The newly accreted gas predicted by the model highly dilutes its ISM, leading to an $\sim1$ dex drop in the metallicity at $t \sim 8$ Gyr. As the gas continues to accumulate, the SF is ignited again, and the metallicity of the stars formed in the second episode can be as low as about $-2$ dex and then rise back up gradually. 

However, the SF enhancement around 8-10 Gyr predicted by the CMD fitting is possibly caused by old blue stragglers rather than by a young small burst. \citet{monelli10a} did not show the derived MEH after $t=6$ Gyr in their result. Therefore, we plot the evolution track after 6 Gyr in gray. When looking at the data covering the entire time range (yellow triangles in panel (d)), it also shows a sudden drop in its MEH, as our model predicts. Although the second SF burst in the SFH of Tucana may not be real, our model demonstrates its ability to predict the observed feature well if such a starburst does exist.

\subsection{Transition-type Dwarf Galaxies}
\label{sec:results_dTran}

Similar to the dSph galaxies, the transition-type galaxies have an old, main episode of SF at an early epoch ($\sim 9$ Gyr ago) followed by a steady decline until $t \approx 5$ Gyr. However, the transition galaxies, LGS 3 and Phoenix, were able to continuously form stars at a low rate until the present day, unlike early-type galaxies whose SF has stopped completely \citep{gallart15}.

The measured SFHs of LGS 3 and Phoenix are shown in the first row of Figure ~\ref{Fig:dtran}. LGS 3 is dominated by a main episode of SF with a maximum occurring about 11.7 Gyr ago and a duration estimated at 1.4 Gyr (FWHM), followed by a rapid decline to a low SFR \citep{hidalgo09}. The SF started in Phoenix at a very early epoch. Phoenix has had ongoing but gradually decreasing SF over nearly a Hubble time. Both galaxies formed 50\% of their total stellar mass about $10.5–11.0$ Gyr ago and more than 95\% of their total stellar mass by 2 Gyr ago \citep{hidalgo11}. 

Based on the continuous SFHs, these two dTran galaxies are predicted to be dominated by gas accretion in the first 2 Gyr (panel (c) in Figure ~\ref{Fig:dtran}), therefore accumulating a large amount of gas (panel (b) in Figure ~\ref{Fig:dtran}). After the first SF episode, which also triggered the outflow, the gas surface density decreased gradually and remained at a low level until the present day. Since SF fluctuated more in Phoenix, the gas exchange process between the galaxy and its external environment might be active (panel (c.4) in Figure ~\ref{Fig:dtran}).

The MEHs of Phoenix and LGS 3 show a similar behavior. They both reached [M/H]$\sim-1.7$ at the earliest time, then enriched by less than 0.1 dex before $t \approx 6$ Gyr. The metallicity started to increase about $\sim$ 7 Gyr ago, after the initial episode of SF.

If no additional galactic wind is assumed in the model (i.e. $\eta=0.0$), the predicted inflow and outflow processes are not significant, so the metallicity will increase very rapidly, as shown by the black dotted lines in panels (f.3) and (f.4) in Figure ~\ref{Fig:dtran}. Again, when $\eta$ increases, stronger outflow wind carries more metal, while the remaining ISM is diluted by more accretion gas, making the enrichment process highly depressed.

Our model does not fit the MEH of Phoenix well in the last 4 Gyr. The CMD-fitted metallicity was 0.5 dex higher than predicted by our model. Considering the SF burst shown between 10 and 12 Gyr in the SFH plot, this overabundance feature might imply an outside origin or a metal-enriched gas infall.

\subsection{Late-type Dwarf Galaxies}
\label{sec:results_dIrr}

Leo A and IC 1613 are the two dIrr galaxies that are still actively forming stars today. Compared to the other types of dwarfs, their metallicities increase continuously and have reached a relatively higher value in the present day. 

Leo A is one of the least luminous gas-rich galaxies we know. It is a small, blue galaxy conspicuous for its population of young, massive stars, with estimated ages ranging from $\sim 10^7$ to $10^8$ yr. The first HST observations of Leo A revealed a very high fraction of low-metallicity stars aged $1 \sim 2$ Gyr \citep{tolstoy98}. Later, \citet{dolphin02} discovered RR Lyrae-type variables hence proving the existence of truly ancient stars in Leo A. Leo A presents the most striking example of delayed SF \citep{cole07}. 

The SFH derived by the CMD fitting method is shown in panel (a) of Figure  \ref{Fig:dIrr} (cyan). Within the ACS/WFC field of view, 90\% of Leo A's SF occurred more recently than 8 Gyr ago. The SFR increased dramatically around $t=5$ Gyr and continued at a high level for 3 Gyr before declining. Another SF boost showed up around 2 Gyr ago. The trigger of SF in the case of Leo A is harder to identify. As suggested by \citet{cole07}, with Leo A being an isolated galaxy for a long time, the gas in it might be present from early on, but only a small fraction of the gas was able to participate in the early SF, with the rest kept warm owing to the possible ultraviolet background or SN feedback heating. The warm gas would have been diffuse and metal-poor, resulting in a long cooling timescale and a possible delay before it could participate in SF  \citep{cole07}. 

Object IC 1613 seems to show a similar trend, albeit to a much less dramatic extent. The SF of this galaxy became more and more intense from $t = 0$ to 3 Gyr, reached a maximum, and then declined \citep{gallart15}. 

Based on the SF law, the increasing SF activity with time implies an increasing amount of cold gas in the reservoir, which is shown in panel (b) of Figure  \ref{Fig:dIrr}. Combining these two histories, our model shows that these two galaxies both receive a continuous net cold gas supplement from their host halos for most of their life, especially in the case of Leo A. 

The minimum gas inflow and outflow are shown as black dashed lines ($\eta=0$) in panels (d) and (e) of Figure  \ref{Fig:dIrr}. When $\eta = 1.0$ is assumed, the strong wind induced by the SF has blown away vast amounts of the enriched gas. The inflow rate evolution history has a similar feature to the outflow rate, except that the peak appears 0.3 Gyr earlier (the time step of the calculation is 0.1 Gyr) compared to the outflow rate. 

The metallicity predicted by our model in the first 8 Gyr is a bit higher than the CMD-fitted value, but the prediction agrees well with observations at the later time. A possible reason for this mismatch is that in the early epoch, the mass of the galaxy was much less than in the present day, so the outflow was more efficient than the model assumed, resulting in a slow enrichment. However, the continued and increasing gas supplement at later times increased the SFR and gradually enriched the ISM. 

Compared to early-type galaxies, the predicted metallicity of Leo A increases much more slowly at the early epoch (panel (f) of Figure  \ref{Fig:dIrr}). In our model prediction, owing to the low SFR at the early time, it took $\sim 2$ Gyr to reach [M/H]$ = -1.5$ dex, much slower than the dSphs. When $\eta = 0.0$, the metallicity would continue to increase after $t = 2$ Gyr and then reach around $-0.3$ dex at the present day, which is much higher than the observed MEH. However, in the case of $\eta = 1.0$, the metallicity was kept at a low level of $-1.5$ dex due to the dilution of the continuously accreted gas and the rejection of the metal-enriched gas. Unlike dSphs, dIrrs contained a large amount of gas all the time and therefore did not show a sudden drop in their MEH, either in the CMD fitting data or in our model prediction. 

\section{Discussion}
\label{sec:dis}

In Section~\ref{sec:results}, we have demonstrated that with only one free parameter, the wind efficiency $\eta$, our SFH-regulated model is capable of predicting a reasonably good MEH for all six LCIDs based on their SFHs. And the model does not require a fine-tuned value for $\eta$: $\eta\sim1.0$ provides good results for all six galaxies. The only difference in the models for these different galaxies is the adoption of different virial velocities, which are determined by their masses. In such a simple and consistent frame, lower-mass galaxies naturally have stronger outflows.

In this section, we compare our new SFH-regulated CEM with the other models in two aspects, namely, the evolution history and the metallicity-versus-gas fraction relation (the $Z - \mu$ relation).

\subsection{Compare to Other CEM Models}
\label{sec:otherCEM}

In this section, we compare our SFH-regulated CEM with two other simple CEM models, i.e., the CB and LKB model. For the bathtub model of \citet{barrera18} (B18), because it assumes a quasi-steady state, no implicit evolution history with time can be presented for individual galaxies (see more discussions in Sec.~\ref{sec:zmu}).

1) In the CB model, it assumes that no gas escapes or flows into the galaxy, i.e. $f_{\text{in}}=f_{\text{out}}=0$. The total mass of the galaxy does not change with time, i.e., $\Sigma_{\text{tot}}(t)\equiv\Sigma_0$, and all baryons are in the gas phase at the beginning time. When a KS law (Equation~(\ref{eq:kslaw})) is taken, the time evolution of the gas, SFR, and metallicity can then be expressed as
\begin{eqnarray}
    \Sigma_{\text{gas}}^{\text{CB}}(t) &=& [(n-1)(1-R)\epsilon~t+\Sigma_0^{1-n}]^{\frac{1}{1-n}} ~,\\
    \Psi^{\text{CB}}(t) &=& \epsilon{\Sigma_{\text{gas}}^{\text{CB}}}^{n}(t) ~,\\
    Z^{\text{CB}}(t)&=& \frac{y}{n-1}\ln\left[\frac{\Sigma_0^{1-n}-(1-n)(1-R)\epsilon~t}{\Sigma_0^{1-n}}\right]~.
\end{eqnarray}
The $\Sigma_0$ is set by assuming that the CB model forms the same stellar mass as the observed one at the present time $t_g=13.5$ Gyr,
\begin{equation}\label{eq:norm}
   \int_0^{t_g} \Psi^{\text{model}}(t)dt = \int_0^{t_g} \Psi^{\text{LCID}}(t)dt~.
\end{equation}

In Figure  \ref{Fig:dsph} - \ref{Fig:dIrr}, we overlaid the predicted $\Psi^{\text{CB}}(t)$ and $Z^{\text{CB}}(t)$ of the CB model for all LCID galaxies by adopting the typical KS law ($\epsilon = 0.25$, $n = 1.4$) and $y = 0.02$ to estimate the global metallicity (blue dotted lines). In the CB model, since the galaxy gathers all of its mass at the beginning and is all gas, the SFR decreases monotonously over time. Obviously, the predicted SFHs do not match the observations, especially for the late-type dwarf galaxies that have increasing SFHs, such as Leo A. The MEHs predicted by the CB model increase very fast over time, resulting in the metallicity of all six LCIDs being more than 1 dex higher than the observed value. 

The CB model may be more suitable for the massive early-type galaxy that gathers all the gas at the beginning and forms stars quickly. It fails in modeling the SF galaxies \citep{barrera18} and the local dwarf galaxies, where the influence of the infall and outflow cannot be ignored.

2) The LKB model, demonstrated by \citet{zhu17}, can fit the local $\Sigma_{*}-\Sigma_\text{gas}/\Sigma_\text{SFR}-Z$ relation obtained from more than 500,000 spatially resolved star-forming regions in disc galaxies observed in the MaNGA survey. The model assumes that the metal-poor gas falls, collapses, and triggers localized SF. There are no radial flows. The SF and metal production are localized within the same region except for the outflowing gas. In this LKB model, the expelled gas should stay within the same area but does not fall back onto the galaxy to participate in subsequent SF. With the above assumptions, the total baryonic surface density, including the stars, the gas that is still in the system, and the gas that has been blown out, remains constant over cosmic time ($\equiv\Sigma_0$),
\begin{equation}
  \begin{split}
    \Sigma_{\text{tot}}(t)\equiv\Sigma_0=\Sigma_*(t)+\Sigma_{\text{gas}}(t)+\Sigma_{\text{out}}(t)~.
  \end{split}
\end{equation}
In the LKB model, the outflow rate is related to the SFR surface density through
\begin{equation}
    f_\text{out}=\frac{d\Sigma_{\text{out}}(t)}{dt}=\eta\Psi(t)~,
\end{equation}
where $\eta$ is the mass loading factor. When a KS law is adopted, the time evolution of each quantity can be expressed as
\begin{eqnarray}
\Sigma_{\text{gas}}^{\text{LKB}}(t) &=& [\Sigma_{0}^{1-n}-\epsilon(1+\eta)(1-n)t]^\frac{1}{1-n}~, \\
\Psi^{\text{LKB}}(t) &=& \epsilon{\Sigma_{\text{gas}}^{\text{LKB}}}^{n}(t)~, \\
Z^{\text{LKB}}(t) &=& Z_0^{\text{LKB}}+\frac{\ln(10)y}{1+\eta}[\log_{10}\Sigma_0-\log_{10}\Sigma_{\text{gas}}^{\text{LKB}}(t)]~.
\end{eqnarray}

\citet{zhu17} assumed that the parameters are all constant ($\epsilon=0.0004$, $n=2.2$, $y_O=0.0026$, $\eta=1.0$, and $Z_0^{\text{LKB}}$=0.1\%$Z_{\odot}$). When an initial gas surface density $\Sigma_{0}$ is given, the LKB model can fully describe the local SFH and chemical evolution. With the most recent data from the MaNGA survey, \citet{zhu17} showed that such a local LKB model can naturally explain the tight relation between stellar surface density, gas surface density, and gas-phase metallicity.

We tested this LKB model on the LCID galaxies and compared the results with the observed SFHs and MEHs. The parameters we adopted were the same as in \citet{zhu17}, except that the yield is scaled to the total yield by assuming $y=2y_O$. The unknown parameter $\Sigma_{0}$ is set by having the same stellar mass as LCID at the present day (Equation~\ref{eq:norm}).
The predicted SFHs and MEHs of the LKB model are plotted as red dashed-dotted lines in Figure  \ref{Fig:dsph} - \ref{Fig:dIrr}.

As shown in Figure  \ref{Fig:dsph} - \ref{Fig:dIrr}, due to the extremely inefficient SF ($\epsilon=0.0004$) and metal production processes \citet{zhu17} assumed in the model, the metallicity grows very slowly, so the MEHs predicted by the LKB model are comparable to observations.
However, the gas is consumed too slowly in the LKB model (panel (b) of Figure  \ref{Fig:dsph} - \ref{Fig:dIrr}), especially for early-type dwarfs. That also prevents the SF process of early-type dwarfs from being quneched at an early stage (panel (a) of Figure \ref{Fig:dsph}). 
In addition, the SFH in the LKB model cannot be fine-tuned, and it decreases monotonously over cosmic time. Therefore, it cannot predict an increasing SFH, which is the case for late-type SF galaxies such as Leo A  (panel (a) of Figure \ref{Fig:dIrr}). 

\subsection{The $Z - \mu$ Relation}
\label{sec:zmu}

\begin{figure*}[!tb]
  \centering
  \includegraphics[width=16.5cm]{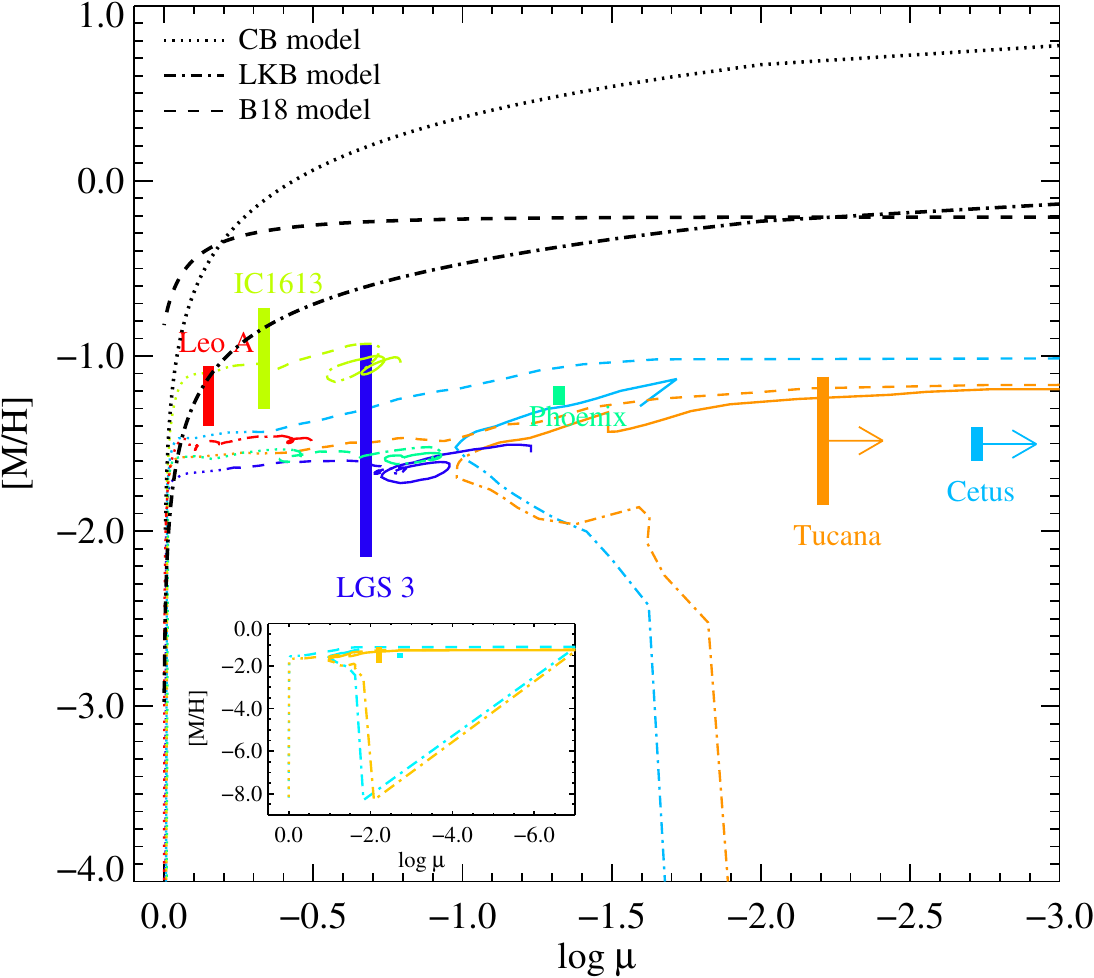}\\
    \caption{Metallicity-gas fraction ($Z - \mu$) relation. The colored vertical bars show the observation data of six LCID galaxies; the gas fractions of LCID galaxies are calculated from the total gas and stellar mass listed in Table \ref{Tab:lcid}, and the metallicity represents the range from the observed metallicity of the region where the CMD was measured to the entire galaxy. The colored lines represent the predictions of our SFH-regulated model for each LCID galaxy with the same color as its observed data. The dotted, dashed, dashed-dotted, and solid lines represent cosmic time ranges of [0, 3], [3, 7], [7, 10], and [10, 13.5] Gyr, respectively. The black dotted, dashed-dotted, and dashed lines represent the evolution track of the CB, LKB \citet{zhu17}, and B18 model \citep{barrera18}, respectively. }
  \label{Fig:zmu}
\end{figure*}

The predictions of CEM models are often used to fit the relations of the observables that can be measured today, such as the SFR, the metallicity $Z$, and the gas fraction $\mu$ which is defined as the gas to total mass ratio. One of the commonly used constraints is the metallicity-versus-gas fraction ($Z - \mu$) relation; that is, galaxies with a lower gas fraction $\mu$ are usually more metal-rich.
In this section, we compare the evolutionary tracks of the $Z - \mu$ relation predicted by different CEMs with observations.

In Figure  \ref{Fig:zmu}, the observed $Z - \mu$ relations of the six LCID galaxies are shown as colored bars. The gas fraction of the whole galaxy $\mu$ is calculated from its total gaseous and stellar mass are listed in column (6) of Table \ref{Tab:lcid}. Since there is no \hi gas detected in Cetus and Tucana, the detection limit of HIPASS \citep{grcevich16} is used to estimate the upper limit of their gas fraction. Note that the gas fraction of the whole galaxy may be different from that of the local region where the CMD was observed. The vertical range of the colored bars shows the possible range of the LCID's metallicity, which is estimated to be between the metallicity measured in the observed region (the last data point of MEH) and the global mean metallicity listed in Table~\ref{Tab:lcid}. Since the observed global mean metallicity estimates from the RGB color or CMD reflect the integral effect of stellar populations, the true [M/H] value may fall in the middle of the colored bar. Therefore, the model prediction is expected to pass through these data bars.

In the case of the CB model, the metallicity is related to the gas fraction in a very simple form,
\begin{equation}\label{eq:CB}
    Z^{\text{CB}} = -y \ln\mu ~,
\end{equation}
where $y$ is the metallicity yield per stellar generation. In Figure  \ref{Fig:zmu}, the black dotted line represents the assumption of $y=0.02$.

In the LKB model, assuming the outflow rate is related to the SFR surface density, the metallicity is then explicitly related to the gas fraction as
\begin{eqnarray}
\label{eq:LKB}
    Z^{\text{LKB}} &=& Z_0^{\text{LKB}} \nonumber \\
    &+& \frac{\ln(10)y}{1+\eta/(1-R)}\cdot\log\left[1+(1+\frac{\eta}{1-R})\frac{1-\mu}{\mu}\right]~,
\end{eqnarray}
where $Z_0^{\text{LKB}}$ is the initial gaseous phase abundance and assumed to be $0.1 \%Z_{\odot}$\citep{zhu17}. We plot the $Z - \mu$ relation track (black dashed-dotted line) predicted by the LKB model in Figure  \ref{Fig:zmu} with the same parameter values as \citet{zhu17} assumed, except that $y$ is doubled in order to calculate the total abundance. 

Following \citet{lilly13}, the gas-regulator model proposed by B18 assumes that the gas mixed with the dark matter flows from the surroundings into a halo. Some fraction of the incoming baryons penetrated down to enter the galaxy system as baryonic gas. There the gas adds to a reservoir within the galaxy. The instantaneous SFR in the galaxy is determined by the instantaneous mass of gas in this internal reservoir. Metals are returned to this internal reservoir. Finally, some gas may be expelled from the reservoir back out to the halo, or even beyond, in a galactic wind. This model is used to explain the local $Z - \mu$ relation observed in more than $9.2\times10^5$ star-forming regions (spaxels) located in 1023 nearby galaxies included in the SDSS-IV MaNGA IFS survey. 

In the B18 gas-regulator model, the SFR is assumed to be linearly proportional to and regulated by the gas mass, $\Psi = \epsilon \Sigma_{\text{gas}}$, and the output rate is proportional to the SFR by $f_{\text{out}} = \eta \Psi$. If this gas-regulator system stays in a quasi-steady state, the time variation of $r_{\text{gas}}$, defined as $\Sigma_{\text{gas}}/\Sigma_*$, could be neglected. The metallicity of the gas-regulated system can be described as 
\begin{equation}
    Z_{\text{eq}}^{\text{B18}} = Z_0 + \frac{y}{1+r_{\text{gas}}+(1-R)^{-1}\left(\eta+\epsilon^{-1}\frac{d\ln(r_{\text{gas}})}{dt}\right)}~,
\end{equation}
where $Z_0$ and $y$ are the initial metallicity and the yield, respectively. This equation explicitly correlates the metallicity with the gas fraction $\mu$ owing to $r_{\text{gas}}=\frac{\mu}{1-\mu}$.
The fitting of this model to the SDSS data results in $\epsilon^{-1}\frac{d\ln(r_{\text{gas}})}{dt}\sim -0.25$. Authors B18 set this factor at this constant value.
We take the same $\eta=1.43$ as B18, but both $Z_0$ and $y$ are doubled because they represent only oxygen in B18. We plot the prediction as the black dashed line in Figure  \ref{Fig:zmu}. 

Our model cannot give an analytical solution for $Z$ to be explicitly related to $\mu$. The equations can be solved numerically only. We have to do the calculation case by case. 
The $Z - \mu$ evolution track predicted by our model has been plotted as the colored line for each LCID in Figure ~\ref{Fig:zmu} based on its own SFH. The dotted, dashed, dashed-dotted, and solid lines representcosmic time ranges of [0, 3], [3, 7], [7, 10], and [10, 13.5] Gyr, respectively. Almost all the tracks have passed through their own observed data bar and stopped at a lower gas fraction. 

It is notable that the $Z - \mu$ evolutionary tracks predicted by our SFH-regulated model are different from all three other models, showing loops in Figure ~\ref{Fig:zmu}. This is caused by the gas cycling process included in our model. When the galaxy consumes gas and forms stars, $\mu$ decreases and $Z$ increases, so the $Z - \mu$ evolution track moves toward the upper right. However, when the fresh primordial gas flows into and feeds the galaxy again, $\mu$ increases and $Z$ decreases, since the ISM has been diluted, causing the track to go to the bottom left. After that, a new episode of SF has been ignited; then $Z$ starts to increase again and gas begins to be consumed. All these processes combine to make a loop on the $Z - \mu$ evolution track.

This phenomenon is more obvious in the dSph galaxies, which are shown in the small panel in Figure ~\ref{Fig:zmu}. The dSph galaxies blow out almost all of the gas after their first SF episode ($t<\sim6$ Gyr), then settle down at an extremely low $\mu$ state. The infall of fresh gas has a significant impact on the $Z - \mu$ evolution track, but the timescale is very short, as also indicated in the MEH in Figure ~\ref{Fig:dsph}.

The feature that there can be multiple values of $Z$ for a given $\mu$ at different evolutionary times of a galaxy cannot be produced by other models. This is obtained naturally when the model is SFH-based. The information on how the gas accumulates, expulsions, or whether it can stay in a quasi-steady state is not assumed but is obtained from the observed SFH.

Our model includes complex gas infall and outflow processes, which could reproduce a nonmonotonic SFH. In reality, the gas infall and outflow processes can be independent events. The SF process can be enhanced by cooling gas accretion or a gas-rich encounter. It can also be depressed due to the SN explosion heating or runoff gas. Therefore, the real SFH should be stochastic, especially in low-mass galaxies.  Our model, which is regulated by the SFH, adapts to a more complicated gas cycling process and hence is more consistent with the large scatters shown by the LCID galaxies in the $Z - \mu$ relation.

\section{Summary}
\label{sec:sum}

In this work, we built an SFH-regulated CEM and tested it on the local LCID galaxies that have precise SFHs and MEHs derived from the deep CMD fitting. The most notable advantage of our model is that the SFH is not subject to any model assumption, so it should be suitable for any shape of SFH. Using the observed SFH as input, we can predict $Z(t)$ for different types of dwarf galaxies in the same frame. The main results are as follows.

(1) Within the same CEM framework, the SFH-regulated model can reproduce reasonable MEHs for all three types of LCID dwarf galaxies based on their observed SFHs.

(2) For dwarf galaxies, the outflow has to occur to reduce their metallicity. The outflow strength of our model is governed by the mass of the galaxies, so it is naturally happens more easily in the lower-mass galaxies because of their shallower potential. Our model can predict reasonable MEHs for all six LCID dwarfs with the same wind efficiency $\eta \sim 1.0$.

(3) Our model can reproduce the  $Z - \mu$ relation of LCIDs. Because of the complex gas infall and outflow processes inferred from the SFH, the evolution track of $Z - \mu$ is nonmonotonic.

Our model is regulated by the SFH, so it can be a useful tool to break the metallicity-age degeneracy in the full spectral fitting. A useful application of our model is to combine it in the full spectral fitting \citep{shen20}. 

The CEM in this paper only considers the global metallicity, as it is the only set of data available. However, the CEM can actually make predictions about the MEHs of different elements, particularly $\alpha$ elements and iron. The metallicity measurements of these elements are crucial for understanding the evolution of galaxies, especially the Milky Way \citep[e.g.][]{Andrews17,Prantzos18,Spitoni21}. Therefore, in the future, obtaining precise measurements of iron and $\alpha$ elements in extragalactic galaxies would significantly enhance the accuracy and usefulness of our CEM.

Owing to the well-known degeneracy between the stellar age and metallicity, different combinations of the single stellar population SSP(age, $Z$) could show a very similar spectrum. The mathematically best-fitting result, such as stochastic SFH and MEH, may not follow the physical principles. Our SFH-regulated CEM could connect the metallicity to the SFH, hence helping to pick up the most probable fitting results.

\begin{acknowledgements}

We thank Evan Skillman and the anonymous referee for their constructive comments that helped to significantly improve the manuscript. This work is supported by the National Key R\&D Program of China (Nos. 2019YFA0405501 and 2022YFF0503402), the Natural Science Foundation of Shanghai (Project No. 22ZR1473000), the National Natural Science Foundation of China (Nos. U1831205, U2031139, 12073059, and 12233005), and the Program of Shanghai Academic Research Leader (No. 22XD1404200). We also acknowledge the science research grants from the China Manned Space Project with No. CMS-CSST-2021-A07.

\end{acknowledgements}

\bibliography{LGevo}{}
\bibliographystyle{aasjournal}

\end{document}